\documentstyle[aps]{revtex}

\begin{document}
\title{Quantum Group SU$_{q}$(2) and\ Pairing in Nuclei}
\author{S. Shelly Sharma\thanks{%
email: shelly@uel.br}}
\address{Departamento de F\'{i}sica, Universidade Estadual de Londrina, Londrina,\\
86051-970, PR, Brazil}
\author{N. K. Sharma}
\address{Departamento de Matem\'{a}tica, Universidade Estadual de Londrina, Londrina,%
\\
86051-970, PR, Brazil}
\maketitle

\begin{abstract}
A scheme for treating the pairing of nucleons in terms of generators of
Quantum Group SU$_{q}$(2) is presented. The possible applications to nucleon
pairs in a single orbit, multishell case, pairing vibrations and
superconducting nuclei are discussed. The formalism for performing BCS
calculations with $q$-deformed nucleon pairs is constructed and the role
played by deformation parameter $q$ analyzed in the context of nucleons in a
single orbit and for Sn Isotopes.
\end{abstract}

Pairing of nucleons manifests itself in the energy Gap in even-even nuclei,
odd-even staggering, moment of inertia of deformed nuclei, low lying 2$^{+}$
states, ground state spins and decay properties of nuclei. Quasi spin
operators, the generators of group SU(2) have been an interesting artefact
for studying nucleon pairing since the time that Racah and Talmi \cite
{raca52} pointed out the group symmetries of the zero range pairing
interaction model. In this talk, a more general scheme for treating the
nuclear pairing problem in terms of generators of quantum group SU$_{q}$(2)
is presented. The quantum group SU$_{q}(2)$, a $q$-deformed version of Lie
algebra SU$(2)$, has been studied extensively \cite{Jimb85,Woro87,Pasq88},
and a $q$-deformed version of quantum harmonic oscillator developed \cite
{Macf89,Bied89}. The quantum group SU$_{q}(2)$ is more general than SU$(2)$
and contains the later as a special case. The underlying idea in using the
zero coupled nucleon pairs with $q$-deformations is that the commutation
relations of nucleon pair creation and destruction operators are modified by
the correlations as such are somewhat different in comparison with those
used in deriving the usual theories. The $q$-deformed theories reduce to the
corresponding usual theories in the limit $q\rightarrow 1$. We first
introduce the seniority scheme and the quasi-spin operators in section I.
The q-deformed nucleon pairs are defined in section II which also contains a
brief review of seniority scheme based on q-deformed nucleon pairs\cite
{she92}.

Section III contains the formulation of random phase approximation (RPA)
with q-deformed nucleon pairs(boson approximation) for nuclei with no
superconducting solution and RPA\ with q-deformed quasi-particle pairs
(quasi-boson approximation) for superconducting nuclei \cite{she94} . In
section IV, the formalism for BCS\ theory with q-deformed nucleon pairs is
presented and it's application to the case of Sn nuclei discussed\cite
{she2000}. The Nucleon pairing in a single $j$ shell has also been treated
by Bonatsos et. al \cite{Bona92,Bona94,Bona99} by associating two $Q$%
-oscillators, one describing the $J=0$ pairs and the other associated with $%
J\neq 0$ pairs. In their formalism, $Q$-oscillators involved reduce to usual
harmonic oscillators as $Q\rightarrow 1$ and the deformation is introduced
in a way different from ours.

\section{Quasi-Spin operators and Seniority Scheme}

Creation and destruction operators for a zero coupled nucleon pair in single
particle orbit $j$\ are, 
\[
Z_{0}=-\frac{1}{\sqrt{2}}\left( A^{j}\times A^{j}\right) ^{0}\hspace{0.3in}%
\text{and}\hspace{0.5in}\overline{Z_{0}}=\frac{1}{\sqrt{2}}\left(
B^{j}\times B^{j}\right) ^{0}
\]
where 
\begin{equation}
A_{jm}=a_{jm}^{\dagger }\text{ ;\quad  }B_{jm}=\left( -1\right)
^{j+m}a_{j,-m}\text{ ;\quad  }a_{jm}^{\dagger }a_{jm}+a_{jm}a_{jm}^{\dagger
}=1.  \label{1}
\end{equation}
With number operator defined as 
\begin{equation}
n_{op}=\sum_{m}a_{jm}^{\dagger }a_{jm}\text{ ,}  \label{2}
\end{equation}
and putting $\Omega =\frac{2j+1}{2}$\ , we can verify that 
\begin{equation}
\left[ Z_{0},\overline{Z_{0}}\right] =\frac{n_{op}}{\Omega }-1;\quad \left[
n_{op},Z_{0}\right] =2Z_{0};\quad \left[ n_{op},\overline{Z_{0}}\right] =-2%
\overline{Z_{0}}\text{ .}  \label{3}
\end{equation}
We can use quasi spin operators defined as

\begin{eqnarray}
S_{j+} &=&\sum_{jm>0}\left( -1\right) ^{j-m}a_{jm}^{\dagger
}a_{j-m}^{\dagger }\qquad ;  \nonumber \\
S_{j-} &=&\sum_{jm>0}\left( -1\right) ^{j-m}a_{j-m}a_{jm}\qquad ,  \label{4}
\end{eqnarray}
and for a single particle orbit $j$\ identify 
\begin{equation}
S_{+}=\sqrt{\Omega }\,Z_{0},\text{ \ \ }S_{-}=\sqrt{\Omega }\,\overline{Z_{0}%
};\hspace{0.2in}S_{0}=\frac{\left( n_{op}-\Omega \right) }{2}.  \label{5}
\end{equation}
The operators $S_{+}$, $S_{-}$\ and $S_{0}$\ are the generators of $SU(2)$\
and satisfy the commutation relations of angular momentum operators, 
\begin{equation}
\left[ S_{+},S_{-}\right] =2S_{0};\hspace{0.2in}\left[ S_{0},S_{\pm }\right]
=\pm S_{\pm }.  \label{6}
\end{equation}
In seniority scheme an $n$\ nucleon state with $v$\ unpaired particles ($v$\
being the seniority) is represented by $\left| n,v\right\rangle $, we have $%
Z_{0}\overline{Z_{0}}\left| v,v\right\rangle =0.$\ The states with $p$\
pairs of nucleons and $n=v+2p$\ can be constructed as 
\[
\left| n,v\right\rangle =N\left( Z_{0}\right) ^{p}\left| v,v\right\rangle .
\]
Choosing a pairing Hamiltonian $H=AZ_{0}\overline{Z_{0}}$\ , the pairing
energy in state $\left| n,v\right\rangle $\ is \cite{law} 
\begin{equation}
E(n,v)=\frac{A}{2\left( 2j+1\right) }\left( n-v\right) \left( 2\Omega
-n-v+2\right) .  \label{6b}
\end{equation}

\section{Nucleon Pairs with q-deformation}

To construct nucleon pairs with q-deformation, we next examine the
generators of quantum group $SU_{q}(2)$. The operators $S_{+}(q)$\ , $%
S_{-}(q)$\ , and $S_{0}(q)$\ satisfy the commutation relations 
\begin{equation}
\left[ S_{+}(q),S_{-}(q)\right] =\left\{ 2S_{0}(q)\right\} _{q};\hspace{0.2in%
}\left[ S_{0}(q),S_{\pm }(q)\right] =\pm S_{\pm }(q),  \label{pair1}
\end{equation}
where 
\begin{equation}
\left\{ x\right\} _{q}=\frac{\left( q^{x}-q^{-x}\right) }{\left(
q-q^{-1}\right) }.  \label{pair2}
\end{equation}
Expressing the creation and annihilation operators for $q-$deformed nucleon
pair as 
\begin{equation}
Z_{0}(q)=\frac{1}{\sqrt{\Omega }}S_{+}(q);\text{\quad \ }\overline{Z_{0}(q)}=%
\frac{1}{\sqrt{\Omega }}\text{\ }S_{-}(q),\hspace{0.2in}  \label{pair3}
\end{equation}
the commutation relations for $q-$deformed nucleon pair creation and
destruction operators are found to be 
\begin{eqnarray}
\left[ Z_{0}(q),\overline{Z_{0}}(q)\right]  &=&\frac{\left\{ n_{op}-\Omega
\right\} _{q}}{\Omega },\quad \left[ n_{op},Z_{0}(q)\right] =2Z_{0}(q); 
\nonumber \\
\text{ \ \ }\left[ n_{op},\overline{Z_{0}}(q)\right]  &=&-2\overline{Z_{0}}%
(q).  \label{pair4}
\end{eqnarray}
The pairing Interaction Hamiltonian is now written as $H(q)=AZ_{0}(q)%
\overline{Z_{0}}(q)$. Using $q=e^{\tau }$, ($\tau \neq 0$) the pairing
energy in seniority scheme for q-deformed pairs is 
\begin{equation}
E_{q}(n,v)==\frac{2A\sinh \left( p\tau \right) \sinh \left[ \left( \Omega
-v-p+1\right) \tau \right] }{\left( 2j+1\right) \sinh ^{2}\left( \tau
\right) }.  \label{pair5}
\end{equation}
Application to various isotopes in single particle orbits $1f_{\frac{7}{2}},$
and $1g_{\frac{9}{2}}$ have shown a good agreement with experimental ground
state energies for small values of deformation parameter\cite{she92}. The
formalism for realizing a multishell calculation was also developed and
applied to Calcium isotopes. It was found that in general weakly interacting
heavily deformed nucleon pairs reproduced the spectra very similar to that
produced by strongly interacting weakly deformed nucleon pairs. However,
depending upon the distance from the closed shell, the energy spectra could
shrink or expand with increase in deformation\cite{she92}.

\section{RPA with q-deformed nucleon pairs and q-deformed Quasi-particle
pairs}

Using the q-deformed pair creation and destruction operators of Eq. (\ref
{pair4}) we derived the Random Phase Approximation equations for the pairing
vibrations of nuclei. For nuclei with no superconducting solution, the boson
creation operator that links the ground state of the nucleus $\left|
A,0\right\rangle $ to the excited eigen state $\nu $ of the $A+2$ nucleon
system with $J^{\pi }=0^{+}$ is defined as 
\begin{equation}
R_{+}^{\nu }=\sum_{m}X_{m}^{\nu }\left( \frac{S_{m+}(q)}{\sqrt{\{\Omega
_{m}\}_{q}}}\right) -\sum_{i}Y_{i}^{\nu }\left( \frac{S_{i+}(q)}{\sqrt{%
\{\Omega _{i}\}_{q}}}\right)  \label{rpa1}
\end{equation}
such that 
\begin{equation}
\left| A+2,\nu \right\rangle =R_{+}^{\nu }\left| A,0\right\rangle \,,\qquad
R_{{}}^{\nu }\left| A,0\right\rangle =0\,.\qquad  \label{rpa2}
\end{equation}
We use the indices $mn(ij)$ for single-particle(hole) levels and $%
R_{{}}^{\nu }=\left( R_{+}^{\nu }\right) ^{\dagger }$.

The equations of motion are set up for $R_{+}^{\nu }$ using single-particle
plus pairing Hamiltonian and the RPA equations for the system obtained using
the commutation relations of eq. (\ref{pair1}). The dispersion relation 
\begin{equation}
\frac{1}{G}=\sum_{n}\frac{\{\Omega _{n}\}_{q}}{(2\epsilon _{n}-\hbar \omega
_{\nu })^{{}}}-\sum_{j}\frac{\{\Omega _{j}\}_{q}}{(2\epsilon _{j}-\hbar
\omega _{\nu })^{{}}}  \label{rpa3}
\end{equation}
along with the nomalization condition easily yields a graphical solution. A
similar procedure is followed for constructing the solution for two-hole
phonon states such that 
\begin{equation}
\left| A-2,\mu \right\rangle =R_{+}^{\mu }\left| A,0\right\rangle \,,\qquad
R_{{}}^{\mu }\left| A,0\right\rangle =0\,,  \label{rpa4}
\end{equation}
where 
\begin{equation}
R_{+}^{\mu }=\sum_{m}X_{m}^{\mu }\left( \frac{S_{m-}(q)}{\sqrt{\{\Omega
_{m}\}_{q}}}\right) -\sum_{i}Y_{i}^{\mu }\left( \frac{S_{i-}(q)}{\sqrt{%
\{\Omega _{i}\}_{q}}}\right) .  \label{rpa5}
\end{equation}
The two phonon states 
\begin{equation}
\left| A,\nu ,\mu \right\rangle =R_{+}^{\nu }R_{+}^{\mu }\left|
A,0\right\rangle  \label{rpa6}
\end{equation}
are the excited $0^{+}$states of the nucleus with excitation energy 
\begin{equation}
E(0^{+})=\hbar \omega _{\nu }+\hbar \omega _{\mu }.  \label{rpa7}
\end{equation}
The q-deformed RPA when applied to study the pairing vibrational states in
the nucleus $^{208}$Pb showed that for $\tau =0.405$ the experimental
excitation energy of the double pairing vibration state and the transfer
cross section for two neutron transfer are well reproduced \cite{she94}.

For superconducting nuclei, one has to construct the quasi-boson creation
and destruction operators from q-deformed quasi-particle pair creation and
annihilation operators. The set of coupled equations 
\begin{eqnarray}
(\hbar \omega _{\nu }-2E_{m})X_{m}^{\nu } &=&-G\sqrt{\{\Omega _{m}\}_{q}}%
\sum_{p}\sqrt{\{\Omega _{p}\}_{q}}\left[ X_{p}^{\nu }\left(
u_{m}^{2}u_{p}^{2}+v_{m}^{2}v_{p}^{2}\right) \right.  \nonumber \\
&&\left. -Y_{p}^{\nu }\left( u_{m}^{2}v_{p}^{2}+v_{m}^{2}u_{p}^{2}\right)
\right]  \label{rpa8}
\end{eqnarray}
and 
\begin{eqnarray}
(\hbar \omega _{\nu }+2E_{m})Y_{m}^{\nu } &=&G\sqrt{\{\Omega _{m}\}_{q}}%
\sum_{p}\sqrt{\{\Omega _{p}\}_{q}}\left[ Y_{p}^{\nu }\left(
u_{m}^{2}u_{p}^{2}+v_{m}^{2}v_{p}^{2}\right) \right.  \nonumber \\
&&\left. -X_{p}^{\nu }\left( u_{m}^{2}v_{p}^{2}+v_{m}^{2}u_{p}^{2}\right)
\right]  \label{rpa9}
\end{eqnarray}
can be solved using standard procedure to furnish the roots $E=\hbar \omega
_{\nu }.$ For testing the formalism, q-deformed boson and quasi boson
approximation calculations for 20 nucleons in two shells were performed. and
compared with exact shell model results. The deformed boson approximation
results for $\tau =i0.104$ and deformed quasiboson approximation energies
for $\tau =0.15$ overlap the exact calculation results in a wide region away
from the phase transition region. The deformation effectively results in
including the correlations left out in normal approximate treatments. One
can expect, therefore, that in a realistic calculation deformation parameter
can be used as a quantitative measure of correlations left out in an
approximate treatment in comparison with the exact results.

\section{ Gap equation in $q$BCS and the Ground State Energy}

Pairing effect can not be interpreted as a contribution to an average static
potential (as in Hartree Fock) or contribution to average vibrating
single-particle potential (as in RPA). It is analogous to Superconductivity
in metals. In 1959 Belyaev\cite{Bely59} successfully applied to nuclei the
Bardeen-Cooper-Schrieffer (BCS) theory originally formulated to explain
superconductivity in metals \cite{BCS}. In view of the usefulness of
formulating the nucleon pairing problem in terms of the generators of SU$_{q}
$(2), we are encouraged to formulate a q-deformed version of BCS theory or
qBCS. Following the idea of building correlations in to the theory by using
pair generators satisfying $q$-commutation relations, we next present the $q$%
-analog of BCS theory ($q$BCS) for nuclei. The formalism when applied to the
case study of $^{114-124}$Sn nuclei elucidates the role played by $q$%
-deformation in these nuclei.

For $N$\ nucleons in $m$\ single particle orbits, we consider the trial wave
function , 
\[
\Psi =\Phi _{j_{1}}\Phi _{j_{2}}...\Phi _{j_{m}} 
\]
where for the orbit $j,$\ 
\begin{equation}
\Phi _{j}=u_{j}^{\Omega _{j}}\sum_{n=0}^{\Omega _{j}}\left( \frac{v_{j}}{%
u_{j}}\right) ^{n}\left[ \frac{\Omega _{j}!}{n!(\Omega _{j}-n)!}\right] ^{%
\frac{1}{2}}\left| n\right\rangle \text{ ;\ }\Omega _{j}=\frac{2j+1}{2}\text{
}  \label{7}
\end{equation}
and 
\[
\text{ }\left| n\right\rangle =\left[ \frac{\left\{ \Omega _{j}-n\right\}
_{q}!}{\left\{ n\right\} _{q}!\left\{ \Omega _{j}\right\} _{q}!}\right] ^{%
\frac{1}{2}}\left( S_{j+}(q)\right) ^{n}\left| 0\right\rangle \text{ } 
\]
is the\ normalized wave function for $n$\ zero coupled nucleon pairs with $q$%
-deformation occupying single particle orbit $j$.

Using a variational approach with the {\bf s}ingle particle plus pairing
Hamiltonian for $q$-deformed pairs given by 
\begin{eqnarray}
H &=&\sum_{r}\varepsilon _{r}n_{op}^{r}-G\sum\limits_{rs}S_{r+}(q)S_{s-}(q)%
\text{ };\;  \nonumber \\
\text{where }r,s &\equiv &j_{1},j_{2},.......j_{m}\text{ ,}  \label{8}
\end{eqnarray}
and the gap parameter defined as, 
\begin{eqnarray}
\Delta (q) &=&G\left\langle \Psi \left| \sum\limits_{r}S_{r+}(q)\right| \Psi
\right\rangle =\sum\limits_{r}\Delta _{r}(q)  \nonumber \\
&=&\sum\limits_{r}Gu_{r}v_{r}\left\{ \Omega _{r}\right\} _{q},  \label{9}
\end{eqnarray}
we obtain{\bf \ }the occupancies, 
\begin{equation}
v_{j}^{2}=0.5\left( 1-\frac{\varepsilon _{j}^{\prime }-\lambda }{\sqrt{%
\left( \varepsilon _{j}^{\prime }-\lambda \right) ^{2}+\left( \Delta (q)%
\frac{\left\{ \Omega _{j}\right\} _{q}}{\Omega _{j}}\right) ^{2}}}\right) ,
\label{10}
\end{equation}
gap parameter 
\begin{equation}
\Delta (q)=\sum_{j}G\,\left\{ \Omega _{j}\right\} _{q}0.5\left( 1-\frac{%
\left( \varepsilon _{j}^{\prime }-\lambda \right) ^{2}}{\left( \varepsilon
_{j}^{\prime }-\lambda \right) ^{2}+\left( \Delta (q)\frac{\left\{ \Omega
_{j}\right\} _{q}}{\Omega _{j}}\right) ^{2}}\right) ^{\frac{1}{2}}
\label{11}
\end{equation}
and consequently\ the gap equation 
\begin{equation}
\frac{G}{2}\sum_{j}\frac{\left\{ \Omega _{j}\right\} _{q}^{2}}{\sqrt{\left(
\varepsilon _{j}^{\prime }-\lambda \right) ^{2}\Omega _{j}^{2}+\left( \Delta
(q)\left\{ \Omega _{j}\right\} _{q}\right) ^{2}}}=1.  \label{12}
\end{equation}
To include the effect of terms containing $u_{j}v_{j}^{3}$\ left out
earlier, we now replace the chemical potential $\lambda $\ by 
\begin{equation}
\lambda (q)=\lambda +\frac{Gv_{j}^{2}\left\{ \Omega _{j}\right\} _{q}}{%
\Omega _{j}}\left( \left\{ \Omega _{j}\right\} _{q}-\Omega _{j}+1\right) .
\label{14}
\end{equation}
The ground state BCS\ energy, $\left\langle \Psi \left| H\right| \Psi
\right\rangle $\ is 
\begin{eqnarray}
E_{bcs}(q) &=&\sum_{j=1}^{m}\left( 2\varepsilon _{j}^{\prime }\,\Omega
_{j}\,v_{j}^{2}-G\,v_{j}^{4}\left\{ \Omega _{j}\right\} _{q}\left( \left\{
\Omega _{j}\right\} _{q}-\Omega _{j}+1\right) \right)  \nonumber \\
&&-\frac{\left( \Delta (q)\right) ^{2}}{G}  \label{15}
\end{eqnarray}
We notice that in a very natural way, the SU$_{q}$(2) symmetry introduces in
the interaction energy, a $q$\ dependence which is linked to the $j$-value
of the orbit occupied by the zero coupled nucleon pairs.

\subsection{Single orbit with 2$\Omega $ degenerate states and Sn nuclei}

For $N$\ nucleons in a single orbit with an occupancy of $2\Omega ,$\ the
ground state wave function is $\Psi =\Phi _{j}$\ and $E_{bcs}(q)$\ is 
\begin{equation}
E_{bcs}(q)=\varepsilon _{j}N-G\left\{ \Omega _{j}\right\} _{q}\frac{N}{%
4\Omega }\left( 2\left\{ \Omega _{j}\right\} _{q}-N+\frac{N}{\Omega }\right)
.  \label{18}
\end{equation}
The exact energy of the $N$\ nucleon zero seniority state,{\bf \ } 
\begin{equation}
E_{exact}=\varepsilon _{j}N-G^{\prime }\frac{N}{4}\left( 2\Omega
_{j}-N+2\right) ,  \label{19}
\end{equation}
can be reproduced ($E_{bcs}(q)=E_{exact})$ by choosing $q$ value and the
pairing strength $G^{\prime }$ such that 
\[
G=\frac{G^{\prime }\Omega _{j}\left( 2\Omega _{j}-N+2\right) }{\left\{
\Omega _{j}\right\} _{q}\left( 2\left\{ \Omega _{j}\right\} _{q}-N+\frac{N}{%
\Omega }\right) }\quad , 
\]
for the choice $\varepsilon _{j}=0.0$ .

In Ref. \cite{she2000} the single orbit limit of $q$BCS is applied to
nuclear sdg major shell with $\Omega =16$, and $4,10,14,20,24,30$ valence
nucleons occupying degenerate $1d_{\frac{5}{2}},0g_{\frac{7}{2}},2s_{\frac{1%
}{2}},1d_{\frac{3}{2}},$and $0h_{\frac{11}{2}}$ orbits. The intensity of
pairing strength required to reproduce $E_{exact}$ decreases with increasing 
$q$ and ultimately $G\rightarrow 0$ for all cases$.$ It is also found that
the strongly coupled zero coupled pairs of BCS\ theory may well be replaced
by weakly coupled $q$-deformed zero coupled pairs of $q$BCS theory. To get
more clues as to whether it is possible to replace the pairing interaction
by a suitable commutation relation between the pairs determined by a
characteristic $q$ value for the system at hand, real nuclei have also been
examined in Ref. \cite{she2000}. There has been an increased interest in the
experimental and theoretical study of Sn isotopes, more so after the
observation of heaviest doubly magic nucleus $_{50}^{100}$Sn$_{50}$ in
nuclear fragmentation reactions \cite{Lewi94,Schn94}. We examined the heavy
Sn isotopes with $N=14,16,18,20,22,$ and $24$ neutrons outside $_{50}^{100}$%
Sn$_{50}$ core. The model space includes $1d_{\frac{5}{2}},0g_{\frac{7}{2}%
},2s_{\frac{1}{2}},1d_{\frac{3}{2}},$ and $0h_{\frac{11}{2}}$, single
particle orbits, with excitation energies $0.0$, $0.22$, $1.90$, $2.20$, and 
$2.80$ MeV respectively. The pairing correlation function $D=\Delta (q)/%
\sqrt{G}$ as a function of $G$ for the cases where deformation parameter
takes some typical successively increasing values varying from $1.0$ to $1.7$
shows some interesting features. In $_{50}^{120}$Sn$_{70}$, pairing
correlations are found to increase as $q$ increases while the pairing
strength $G$ is kept fixed. For $q=1.0$ that is conventional BCS theory the
pairing correlation vanishes for $G<G_{c}(\sim 0.065$ MeV$)$ as expected. We
find $D$ going to zero for successively lower values of coupling strength,
for example $G_{c}\sim 0.04$ MeV for $q=1.3$ as the deformation $q$ of zero
coupled pairs increases. We may infer that the $q$BCS takes us beyond BCS
theory.

The sets of $G,q$ values that reproduce the empirical $\Delta $ for $%
_{50}^{120}$Sn$_{70}$ are next used to calculate the gap parameter $\Delta $
and the ground state BCS energy E$_{N},$ for even isotopes $^{114-124}$Sn
and compared with the experimental values of $\Delta $ in Ref.\cite{she2000}%
. The results of $q$BCS for Sn isotopes are not much different from BCS as
far as the Gap parameter $\Delta $ is concerned. The ground state binding
energies are however lowered by the deformation. The pairing correlations,
measured by $D=\Delta (q)/\sqrt{G}$, are seen to increase as $q$ increases
(for $q$ real) while the pairing strength $G$ is kept fixed, in Sn isotopes.
It is immediately seen that $q$ parameter is a very good measure of the
pairing correlations left out in the conventional BCS theory. The underlying 
$q$-deformed nucleon pairs show increasingly strong binding as the value of $%
q$ is increased. It opens the possibility of obtaining the exact correlation
energies by choosing appropriately the combination of $G,q$ values.

The results of our study of qBCS are consistent with our earlier conclusions 
\cite{she92,she94} that the $q$-deformed pairs with $q>1$ $(q$ real) are
more strongly bound than the pairs with zero deformation and the binding
energy increases with increase in the value of parameter $q$. In contrast by
using complex $q$ values one can construct zero coupled deformed pairs with
lower binding energy in comparison with the no deformation zero coupled
nucleon pairs\cite{she94}. In general the pairing correlations in $N$
nucleon system, measured by $D=\Delta (q)/\sqrt{G}$ , increase with
increasing $q$ (for $q$ real) and $q$BCS takes us beyond the BCS theory. The
formalism can be tested for several other systems, for example metal grains,
where cooper pairing plays an important role.

\section{Acknowledgments}

S. Shelly Sharma and N. K. Sharma acknowledge support from Universidade
Estadual de Londrina.

\end{document}